\documentclass[10pt]{bmc_article}

\usepackage{cite} 
\usepackage{url}  
\usepackage{ifthen}  
\usepackage{multicol}   
\usepackage[utf8]{inputenc} 
\usepackage{amssymb,amsmath,graphicx}
\usepackage{graphicx}
\usepackage{graphics}
\usepackage{lscape}

\newboolean{publ}

\newenvironment{bmcformat}{\baselineskip20pt\sloppy\setboolean{publ}{false}}{\baselineskip20pt\sloppy}

\begin{document}
\begin{bmcformat}


\title{Complex-based analysis of dysregulated cellular processes in cancer}


\author{Sriganesh Srihari\correspondingauthor$^1$,%
         \email{SS1\correspondingauthor -- s.srihari@uq.edu.au}       
         Piyush B. Madhamshettiwar$^1$,%
         \email{PBM -- p.madhamshettiwar@uq.edu.au}         
         Sarah Song$^1$,%
         \email{SS2 -- sarah.song@uq.edu.au}                  
         Chao Liu$^1$,%
         \email{CL -- chao.liu@imb.uq.edu.au}         
         Peter T. Simpson$^2$,%
         \email{PTS -- p.simpson@uq.edu.au}         
         Kum Kum Khanna$^3$,%
         \email{KKK -- kumkum.khanna@qimrbergofer.edu.au}
       and
       	 Mark A. Ragan\correspondingauthor$^1$%
       	 \email{MAR\correspondingauthor -- m.ragan@uq.edu.au}%
      }


\address{%
    \iid(1)Institute for Molecular Bioscience, The University of Queensland, St. Lucia, Queensland 4072, Australia.
    \iid(2)The University of Queensland Centre for Clinical Research, Brisbane, Queensland 4006, Australia.
    \iid(3)QIMR-Berghofer Institute of Medical Research, Brisbane, Queensland 4006, Australia. %
}%

\maketitle


\begin{abstract}
   	\noindent\textbf{Background:}
   	Differential expression analysis of (individual) genes is often used to study their roles in diseases.   	
   	However, diseases such as cancer are a result of the combined effect of multiple genes. Gene products such as proteins   	
   	seldom act in isolation, but instead constitute stable multi-protein complexes performing dedicated functions.
   	Therefore, complexes aggregate the effect of individual genes (proteins) and can be used to gain a better understanding of cancer mechanisms.
   	Here, we observe that complexes show considerable changes in their expression, in turn directed by the concerted action of transcription factors (TFs), across cancer conditions.
   	We seek to gain novel insights into cancer mechanisms through a systematic analysis of complexes and their transcriptional regulation.

        \noindent\textbf{Results:}        
        We integrated large-scale protein-interaction (PPI) and gene-expression datasets to identify complexes
        that exhibit significant changes in their expression across different conditions in cancer. 
        We  devised a \emph{log-linear model} to relate these changes to the differential regulation of complexes by TFs.
        The application of our model on two case studies involving pancreatic and familial breast tumour conditions revealed:               
        (i) complexes in core cellular processes, especially those responsible for maintaining genome stability and cell proliferation (\emph{e.g.} DNA damage repair and cell cycle) show considerable changes in expression;
        (ii) these changes include decrease and countering increase for different sets of complexes indicative of \emph{compensatory mechanisms} coming into play in tumours; and
        (iii) TFs work in cooperative and counteractive ways to regulate these mechanisms. 
        Such aberrant complexes and their regulating TFs play vital roles in the initiation and progression of cancer.

       	\noindent\textbf{Conclusions:}
       	Complexes in core cellular processes display considerable decreases and countering increases in expression, strongly reflective of compensatory mechanisms in cancer.
       	These changes are directed by the concerted action of cooperative and counteractive TFs. 
       	Our study highlights the roles of these complexes and TFs and presents several case studies of compensatory processes, thus providing novel insights into cancer mechanisms.
        Availability: \url{http://www.bioinformatics.org.au/tools-data} -- CONTOUR\emph{v}2

\end{abstract}

\ifthenelse{\boolean{publ}}{\begin{multicols}{2}}{}


\section*{Background}

Transcriptional regulation is a fundamental mechanism by which all cellular systems mediate the activation or repression of genes, 
thereby setting up striking patterns of gene expression across diverse cellular conditions -- 
\emph{e.g.} across cell-cycle phases~\cite{Elkon2003, Srihari2012a,Bar-Joseph2007}, normal \emph{vs} cancer states~\cite{Bar-Joseph2007} or stress conditions~\cite{Spriggs2010}.
Such regulation of gene expression is executed by the concerted action of
transcription factors (TFs) that bind to specific regulatory DNA sequences associated with target genes~\cite{Luscombe2004,Balaji2006}.
Deciphering the roles of TFs is a significant challenge and has been the focus of numerous studies,
with great interest being recently shown in cancer~\cite{Spriggs2010,Bar-Joseph2007,Nebert2002,Darnell2002,Karamouzis2011}.
For example, Bar-Joseph \emph{et al.}~\cite{Bar-Joseph2007}
identified periodically expressed cell-cycle genes in human foreskin fibroblasts to understand their differential regulation between normal and cancer conditions.
Nebert~\cite{Nebert2002} surveyed TF activities in cancer, emphasizing the roles of TFs as
proto-oncogenes (gain-of-function) that serve as accelerators to activate the cell cycle, and as tumour suppressors (loss-of-function) that serve as brakes to slow the growth of cancer cells.
Darnell~\cite{Darnell2002} classified TFs having cancerous or oncogenic potential into three main kinds -- steroid receptors (\emph{e.g.} oestrogen receptors in breast cancer and androgen receptors in prostate cancer),
resident nuclear proteins activated by serine kinase cascades (\emph{e.g.} JUN and FOS), and 
latent cytoplasmic factors normally activated by receptor-ligand interaction at the cell surface (\emph{e.g.} STATs and NF$\kappa$B).
Darnell~\cite{Darnell2002} also discussed the signalling pathways of these TFs (including Wnt-$\beta$-catenin, Notch and Hedgehog signalling) as potential drug targets in cancer.
Karamouzis and Papavassiliou~\cite{Karamouzis2011} discussed rewiring of transcriptional regulatory networks in breast tumours focusing on subnetworks of 
estrogen receptor (ERs) and epidermal growth factor receptor (EGFRs) family members.

Most studies focus on transcriptional regulation of individual target genes. However, diseases such as cancer are a result of the combined effect of multiple
genes. Gene products such as proteins seldom act in isolation, but instead physically interact to constitute \emph{complexes} that perform specialized functions.
Studying protein complexes therefore provides an aggregative or ``systems level" view of gene function and regulation than studying individual proteins (genes).
Here we integrate large-scale protein-interaction (PPI) and gene-expression datasets to examine the differential regulation of complexes across cancer conditions.

\subsection*{An initial analysis}
We compiled a list of protein complexes by clustering a network of human PPIs.
Co-functional (interacting) proteins are encoded by genes showing high mRNA co-expression~\cite{Jansen2002,Grigoriev2001}. Therefore, we
quantified the ``functional activity" for each of these complexes by aggregating pairwise co-expression values between their constituent proteins.
Analysis for two pancreatic-tissue conditions \emph{viz} normal and ductal adenocarcinoma (PDAC) tumour revealed 
significant changes in co-expression for these complexes between the two conditions.
For example (Figure 1), CHUK-ERC1-IKBKB-IKBKG showed a change in co-expression, interestingly
coinciding with changes in its transcriptional regulation by the NF$\kappa$B-family of TFs.
This complex constitutes the serine/threonine kinase family, while the TFs play essential roles in NF$\kappa$B signalling pathway
(\url{www.genecards.org})~\cite{SafranM2002}, which are implicated in PDAC~\cite{Jones2008,Fujioka2003}.

Based on these observations, here we seek to understand differential co-expression of complexes and its relationship with differential regulation by TFs between cancer conditions.
Therefore we:
\begin{itemize}
\item devise a computational model to identify complexes showing significant differential co-expression and the TFs regulating these complexes; and
\item apply the model on two case studies -- normal \emph{vs} PDAC tumour and  BRCA1 \emph{vs} BRCA2 familial breast tumour conditions -- to decipher their roles in these tumours.
\end{itemize}
In summary (see Methods for details) we compute co-expression values for each of the complexes under different cancer conditions.
We then introduce a log-linear model to relate changes in co-expression of complexes to changes in their regulation by TFs between these conditions.
We apply the model to identify influential TFs and complexes and validate their roles in cancer.


\section*{Results}

\subsection*{Experimental datasets}

\noindent\emph{PPI data:}
We gathered \emph{Homo sapiens} PPIs identified from multiple low- and high-throughput experiments deposited in 
Biogrid (v3.1.93)~\cite{Stark2010} and HPRD (2009 update)~\cite{Keshava2009}.
To minimize false positives in these PPI datasets, we employed as scoring scheme Iterative-CD~\cite{Liu2009} (with 40 iterations)
to assign a reliability score (between 0 and 1) to every interaction, and then discarded all low-scoring interactions ($<0.20$) to build a
dense high-quality PPI network of $29600$ interactions among $5824$ proteins (average node degree 10.16).
\vspace{1.5mm}

\noindent\emph{Gene expression data:}
We have performed one of the largest gene expression profiling of familial breast tumours ($n=74$) and stratified them based on BRCA mutation status as
BRCA1-, BRCA2- and non-BRCA1/2 tumours~\cite{Waddell2010}. Among these,
BRCA1 and BRCA2 tumours are phenotypically most different~\cite{Lakhani1998} and we consider these two for our analysis here;
our dataset contains 19 BRCA1 and 30 BRCA2 expression samples (GEO accession GSE19177).
In addition, we also gathered expression samples from pancreatic tumours -- 
normal and PDAC matched ($39$ in each) -- from the Badea \emph{et al.} study~\cite{Badea2008} (GSE15471).

Sporadic breast tumours constitute 93-95\% of all breast tumours and most studies classify these into the four molecular subtypes,
luminal-A, luminal-B, basal-like and HER2-enriched~\cite{Perou2000,Sorlie2001, Taherian-Fard2014}.
Broadly, basal-like tumours do not express the ER, PR and HER2 receptors, and exhibit
high aggressiveness and poor survival attributed to distant metastasis, compared to luminal tumours.
However, much less is known about familial tumours (the remaining 5-7\%), although studies~\cite{Waddell2010,Lakhani1998,Taherian-Fard2014} have noted that BRCA1 tumours are predominantly basal-like while
BRCA2 tumours are more hetergeneous and may be HER2-enriched or luminal-like.

Pancreatic tumours, on the other hand, are more uniform with PDAC accounting for most ($95$\%) pancreatic tumours and is predominantly characterized by 
dysfunctioning (by mutation) of the KRAS oncogene and of the CDKN2A, SMAD4 and TP53 tumour-suppressor genes~\cite{Jones2008}.

\noindent\emph{Transcription factors:}
We gathered $1391$ TFs from Vaquerizas \emph{et al}.~\cite{Vaquerizas2009}, manually curated from a combined assessment of DNA-binding capabilities, evolutionary conservation and
integration of multiple sources.

\noindent\emph{Benchmark complexes:}
For independent validation, we used manually curated human complexes from CORUM~\cite{Ruepp2009}, a total of 1843 complexes of which we used 722 having size at least 4.

\noindent\emph{Benchmark genes and TFs in cancer:}
For validation we used known (mutation-driver) genes (total 118) from COSMIC~\cite{Bamford2004} and known TFs (total 82) in cancer from~\cite{Patel2013}.

\subsection*{Analysis of PPI networks highlights considerable rewiring between tumour conditions}

By integrating PPI and gene expression datasets (see Methods)
we obtained two pairs of conditional PPI networks -- normal-PDAC for pancreatic and BRCA1-BRCA2 for breast tumours. 
Figure 2 shows the co-expression-wise distribution for protein pairs in these networks.
Normal \emph{vs} PDAC displayed striking differences in these distributions
(KS test: $D_{NP} = 23.11 > K_{\alpha = 0.05} = 1.36$), reflecting considerable rewiring of PPIs.
PDAC showed significant loss in co-expression for both positively co-expressed as well as
negatively co-expressed interactions compared to normal, indicative of both disruption as well as emergence of interactions in the tumour. 
Such rewiring has also been noted in earlier studies~\cite{Chu2008,Srihari2014}.

Strikingly enough BRCA1 \emph{vs} BRCA2 tumours also showed significant differences in PPI distributions (Figure 2)
(KS test: $D_{B12} = 22.85 > K_{\alpha = 0.05} = 1.36$), reflecting considerable differences in PPI wiring between the two breast tumours.
BRCA1 tumours displayed higher co-expression compared to BRCA2 tumours, $\sim$15700 PPIs with higher correlations.

DAVID-based~\cite{Dennis2003} functional analysis of these rewired interactions ($\Delta \geq0.50$)
showed significant ($p \leq 0.001$) enrichment for the \emph{Biological Process} (BP) terms --
Cell cycle, Chromatin organisation, DNA repair and RNA splicing, indicating considerable rewiring in core cellular processes responsible for genome stability.
Among these were interactions involving the tumour suppressors TP53 and SMAD4 in PDAC, which are known genes mutated in the tumour, and
the DNA double-strand break (DSB) repair proteins BRE and BRCC3 along with BRCA1, BRCA2 and TP53, in breast tumours.

\subsection*{Analysis of complexes highlights disruption to core cellular mechanisms in tumours}

Matching of complexes using $t_J = 0.67$ and $\delta = 0.10$ (Methods) resulted in a total of 256 and 277 matched complexes ($\mathcal{M})$ for normal-PDAC and BRCA1-BRCA2 conditions, respectively
(Table 1). The co-expression-wise distributions (Figure 3) revealed significant differences for both normal \emph{vs} PDAC as well as BRCA1 \emph{vs} BRCA2 conditions
(KS test: $D_{NP} = 1.69 > K_{\alpha = 0.05} = 1.36$ in pancreatic and $D_{B12} = 5.48 > K_{\alpha = 0.05} = 1.36$ in breast),
indicating that rewiring in PPI networks had considerable impact on these complexes.
Overall, we noticed considerable drop in co-expression for PDAC \emph{vis-a-vis} normal, whereas
BRCA1 tumours showed higher co-expression \emph{vis-a-vis} BRCA2 tumours (Figures 3 and 4).
These differences were larger towards the higher co-expression ranges which correspond better to active complexes (Figure 3), indicating that cellular functions were considerably impacted in these tumours.
These observations were reproducible using an independent set of complexes from CORUM (Figures 3 and 4) and 
were significantly ($p<0.001$) greater than expected by random (using 500 random complexes generated 1000 times).

DAVID-based analysis for complexes displaying changes $\geq 0.4$ indicated significant ($p<0.001$) enrichment for core cellular pathways involved in genome stability including Cell cycle and DNA repair (Table 2).
The complexes in PDAC were enriched for TGF-$\beta$, Wnt and NF$\kappa$B signalling, all of which are
implicated in pancreatic cancer~\cite{Jones2008,Biankin2009,Zeng2006,Biankin2012}.
The complexes in breast tumours reflected aberration in Homologous recombination (HR), a key DSB-repair pathway which includes the breast cancer susceptibility genes BRCA1 and BRCA2.

\subsection*{Analysis of complexes reveal compensatory mechanisms activated in tumours}
We next divided the set of matched complexes $\mathcal{M}$ into two subsets:
\begin{itemize}
\item $\mathcal{M}'$ -- those with \emph{higher} co-expression in normal \emph{vis-a-vis} PDAC, or \emph{higher} co-expression in BRCA1 tumours \emph{vis-a-vis} BRCA2 tumours; and
\item $\mathcal{M}''$ -- those with \emph{lower} co-expression in normal \emph{vis-a-vis} PDAC, or \emph{lower} co-expression in BRCA1 tumours \emph{vis-a-vis} BRCA2 tumours.
\end{itemize}
Table 3 shows changes in co-expression ($\Delta C$) observed for $\mathcal{M}'$ and $\mathcal{M}''$. 
While most complexes showed a decrease in co-expression from normal to PDAC (159 out of 256) and from BRCA1 to BRCA2 tumours (225 out of 277),
interestingly a considerable number of complexes showed an increase (96 and 52). 
But, the decrease was steeper compared to the increase (max: 0.969 \emph{vs} 0.421 and 0.761 \emph{vs} 0.543; avg: 0.336 \emph{vs} 0.192 and 0.281 \emph{vs} 0.197).
Similar trends were observed using CORUM complexes and were significantly ($p<0.001$) greater than expected by random.
We suspect these observations are indicative of \emph{compensatory mechanisms} coming into play in these tumours, as explained below.
\vspace{1.5mm}

In the classical work on ``hallmarks of cancer", Hanahan and Weinberg~\cite{Hanahan2000} describe seven to ten key distinguishing hallmarks of tumour cells, among which are limitless replicative potential
and self-sufficiency in growth signals. Cellular mechanisms including cell cycle and DDR are considerably weakened in tumour cells, but these cells survive on
last-standing mechanisms (weak links) to continue proliferation. This is due to the activation of compensatory or back-up mechanisms.
Although these compensatory mechanisms cannot completely substitute for the weakened or disrupted ones,
these are sufficient to enhance the survival of tumour cells~\cite{Hanahan2000,Logue2012}.
Our analysis reflect such compensatory trends -- a fraction of complexes showed increase in co-expression, but the increase was not as steep as the decrease for the remaining faction.
However, a straightforward Gene Ontology analysis is too general to delineate the roles of the two factions because both originate from the same or similar processes. We therefore 
investigated a few specific cases (below).

\subsubsection*{Examples of compensatory mechanisms and validation for roles in cancer}

\paragraph{Normal \emph{vs} PDAC tumour} (Figure 5a):
DSB-repair functionality is severely impacted in PDAC~\cite{Wang2008,Li2006},
with inactivating mutations in RAD50 and NBS1 attributed to loss of DSB-repair functionality increasing the risk of pancreatic cancer~\cite{Wang2008}.
DSBs are detected by the MRE11-RAD50-NBS1 (MRN) and Ku70/Ku80 (XRCC6/XRCC5) complexes in the HR and non-homologous end-joining (NHEJ) pathways, respectively.
In HR, the repair process involves recruitment of the BRCA1-A complex (BRCA1-BARD1-FAM175A-UIMC1-BRE-BRCC3-MERIT40) to sites of DSBs.
We observed a decrease in co-expression for all the three complexes, indicating considerable weakening of the DSB machinery.
On the other hand, we noticed an increase in co-expression for the single-strand break (SSB) and mismatch (MMR) repair complexes XRCC1-POLB-PNKP-LIG3 and MSH6-MLH1-MLH2-PSM2-PCNA, respectively.
The XRCC1 complex is responsible for SSB repair through sister chromatid exchange following DNA damage by ionizing radiation, while the MSH6 complex is involved in the recognition and repair of
mispairs. Together these observations suggest the activation of SSB and MMR machinery compensating for the loss in DSB-repair machinery;
such a functional relationship has been observed previously between DSB and SSB repair pathways~\cite{Gottipati2010}. 
\vspace{1.0mm}

The NF$\kappa$B signalling pathway has been strongly implicated in KRAS signalling and pancreatic tumorigenesis~\cite{Campbell1998,Koorstra2008}. Consistent with this, we noticed considerable
changes in co-expression for several NF$\kappa$B complexes including the NF$\kappa$B1/REL family, which plays important roles in programmed cell death and proliferation control and is critical in 
tumour initiation and progression~\cite{Koorstra2008}.
The calcium-binding proteins S100A2, S100A8 and S100A9 are known to modulate P53 activity~\cite{Mueller2005}
and their over-expression has been associated with metastatic phenotype of pancreatic cancer~\cite{Biankin2009}.
The inactivation of the RAS-associated RASSF1A and RASSF5 complexes, which act as tumour suppressors~\cite{Dammann2003,ParkJ2010}, is
frequent in pancreatic cancer~\cite{Dammann2003}.
The complex DDX20-GEMIN4-PPP4C-PPP4R2 associated with the SMN (survival of motor neuron), and
SNAP23-STX4-VAMP3-VAMP8 associated with vesicular transport, docking and/or fusion of synaptic vesicles with the presynaptic membrane (\url{www.genecards.org})~\cite{SafranM2002},
support tumorigenic invasion of neural cells in pancreatic cancer~\cite{Biankin2012}.
\vspace{1.5mm}

\paragraph{BRCA1 \emph{vs} BRCA2 tumours} (Figure 5b):
We observed a lower co-expression for the MMR complex MLH1-MSH6-MSH2-PMS2-PCNA in BRCA1 tumours compared to BRCA2 tumours;
we think this is due to the parallel roles of BRCA1.
BRCA1 has a key role in DSB repair, and BRCA1-deficient cells have defects in the two DSB repair pathways HR and NHEJ~\cite{LiuC2014}.
BRCA1 associates with PCNA and the mismatch repair proteins MSH2, MSH6 and MLH1 to form the BASC complex, a genome-surveillance complex required to sense and repair DNA damages~\cite{WangY2000},
thereby also playing a role in the MMR pathway.
On the other hand, BRCA2 has been associated with functions only in HR~\cite{Khanna2001,Zhuang2006,Thompson2011,JiangG2013}.
Therefore, we suspect that although MMR pathway is compensatorily activated in response to DSB-repair deficiency, BRCA1 tumours exhibit a weaker MMR pathway compared to BRCA2 tumours because of the direct
involvement of BRCA1 in the MMR pathway.
\vspace{1.0mm}

The DSS1 complex consisting of BRCA2, DSS1 and the integrator subunits mediates the 3'-end processing of small nuclear RNAs~\cite{BaillatD2005}, and
BRCA2 deficiency could result in a reduced stability of this complex.
The expression of replication factor C complex (RFC2, RFC3 and RFC4) is indicative of proliferative potential (high cell division rates) of BRCA1 tumours. We noticed over-expression of this complex
in BRCA1 compared to BRCA2 tumours.

Finally, a considerable number of cancer genes from COSMIC Classic were represented in complexes showing changes $\Delta C \geq 0.10$ (Figure 6), suggesting that
differential co-expression of complexes is a strong indicator of tumorigenic processes.


\subsection*{Relating changes in co-expression complexes to their transcriptional regulation}
We computed Pearson and Spearman rank coefficients between changes in co-expression of complexes and their transcriptional regulation as follows.
For each complex-pair $\{S_s,T_t\} \in \mathcal{M}(\mathcal{S},\mathcal{T})$, we measured its change in correlation $\Delta C(S_s,T_t)$,
and the total change in its regulation by TFs $F_f$, $ \sum \Delta R = \sum_{f=1}^k \Delta R((S_s,T_t),F_f)$ (see Methods).
This resulted in $226$ complex-TF pairs in pancreatic and $241$ in breast with non-zero $\Delta C$ and $\Delta R$.
Note that we lose at most 13\% of complexes (pancreatic: 256 down to 226, breast: 277 down to 241) as a result of our requirement that TFs interact with at least one complexed protein
(Methods).
We observed positive Pearson and Spearman coefficients which were supported by CORUM complexes (Table 4).
The Spearman coefficients were higher than Pearson in both cases, indicating a non-linear relationship; this supports our use of a log-linear model (Methods).

\subsection*{Analysing influential TFs in pancreatic and breast tumours}
Table 5 lists the TFs with non-zero \emph{overall} influence identified using our model (see Methods).
Extrapolating from the simplified example (see Methods), the $+$ and $-$ signs can be interpreted as
cooperative and counteractive action of TFs in regulating complexes. As these are overall influence values (that is, across all complexes and TFs), it is difficult to
interpret this straightaway. Therefore, we restrict our focus to only STAT1 and STAT3.
These two TFs are directly involved in pancreatic tumorigenesis and proliferation, and are thought to play opposite roles -- while STAT1 promotes apoptosis, STAT3 is
essential for the proliferation and survival of tumour cells~\cite{Pensa2009}. Solving Equation~\ref{eq:infl3} for STAT1 and STAT3 using only the subset of complexes
they share (\#90), we obtained $\gamma$(STAT1) = $1.714$ and $\gamma$(STAT3) = $- 1.582$, \emph{i.e.} these are counteractive TFs (Methods).
Their shared complexes were enriched for Cell cycle, Apoptosis and RAS signalling,
consistent with the counteractive roles for STAT1 and STAT3~\cite{Pensa2009}.

Differential expression analysis using limma~\cite{Smyth2004} for normal \emph{vs} PDAC indicated that most of the influential TFs were significantly up- or down-regulated (Table 5).
But, a few influential TFs did not show such differential expression, for example heat shock factor-1 (HSF1).
Investigation into the complexes regulated by HSF1 revealed considerable changes in co-expression for the cysteine-aspartic acid protease (caspase) family including
CASP10-CASP8-FADD-FAS (from $1.28$ to $-0.019$), documented in CORUM~\cite{Ruepp2009} under the functional category `40.10.02: Apoptosis'.
Caspases are involved in signal transduction pathways of apoptosis, necrosis and inflammation (\url{www.genecards.org})~\cite{SafranM2002}, and the role of HSF1 in
regulating caspases thereby contributing to the pathogenesis of pancreatic cancer has been investigated~\cite{Dudeja2011}.

In the case of BRCA1 \emph{vs} BRCA2 tumours, only four of the influential TFs (GATA3, ESR1, FOXA1 and XBP1) were identified as differentially expressed.
These four TFs are ER targets. BRCA1 tumours being predominantly basal-like, do not express ER and therefore show lower expression of ER targets compared to
BRCA2 tumours, which are predominantly luminal-like and express ER~\cite{Lakhani1998}.
Additionally, Joshi \emph{et al.}~\cite{JoshiH2012}, using a pathway-based analysis, have noted over-representation of ESR1, GATA3, MYC, XBP1, FOXA1 and MSX2 in luminal tumours, and
NF$\kappa$B1, C/EBP$\beta$, FOXO3, JUN, POU2F3 and FOXO1 in basal-like tumours.
We also found higher expression of the NF$\kappa$B-signalling TFs in BRCA1 tumours --
the complex NF$\kappa$B1-NF$\kappa$B2-REL-RELA-RELB composed entirely of NF$\kappa$B TFs, showed a higher correlation in BRCA1 tumours than BRCA2 tumours.
This is consistent with earlier findings~\cite{JoshiH2012,BiswasDK2004} that ER-negative tumours (BRCA1 tumours) display aberrant expression of NF$\kappa$B which makes these
tumours highly aggressive.
\vspace{1.5mm}

These observations also suggest that differential expression is not sensitive enough to identify all the genes (here, TFs) involved in tumours.
Many of the TFs may not be differentially expressed themselves, but are differentially
\emph{co-expressed} with their target genes. One such possible situation occurs when the TFs themselves are not mutated or (epigenetically) silenced, but their target genes are.

Finally, $12$ of $37$ TFs in pancreatic, and $14$ of $40$ TFs in breast tumours were among the $82$ cancer TFs listed in~\cite{Patel2013}.
DAVID-based functional analysis of TFs showed significant enrichment for several pathways in cancer ($p<$1.1E-05, 23.1\% genes), in particular the JAK-STAT pathway ($p<$1.9E-02, 10.3\% genes), a
known driver pathway in cancer~\cite{Pensa2009}.

\section*{Discussion}

We had observed considerable PPI rewiring \emph{via} differential co-expression analysis (Figure 2).
In Figure 7, we now show the PPI network for normal \emph{vs} PDAC with interactions weighted by the differential co-expression values.
Figure 7a highlights the largest component (558 proteins and 519 interactions), which shows an overall decrease in co-expression. A considerable number of genes in this component are
targets of ubiquitination (UBC) and sumoylation (SUMO1 and SUMO2) (Figure 7b) possibly causing their inactivation. 
However, there are several pockets showing increase in co-expression. Interestingly some of the
genes topologically central to these pockets are known drug targets in PDAC (Figure 7c), \emph{e.g.} PLK1~\cite{Zhang2012} and ANAX2~\cite{Zheng2012}.
Similarly PELP1 which interacts directly with STAT3 and is responsible for cell proliferation and survival in several tumours~\cite{Pensa2009}, is likewise an identified drug target in PDAC~\cite{Kashiwaya2010}.
A similar analysis in BRCA1 \emph{vs} BRCA2 tumours highlighted increase in PPI co-expression around the mitotic regulators CDK1, CDC20 and CKS1B and the histone deacetylases HDAC1 and HDAC6; 
these are known drug targets for which inhibitors have been developed~\cite{Sutherland2009,Marks2000}.

We clustered this normal \emph{vs} PDAC network using MCL (inflation 2.3) both with and without the weights as input, and we observed that most clusters 
predominantly constitute only one kind of interactions, either those showing increase or decrease in co-expression --
of the 30 clusters of sizes $\geq4$, 17 had at least two-third interactions showing decrease, 9 had at least two-third interactions showing increase.
Among these, PLK1 belonged to a cluster in which all interactions showed an increase (Figure 7d). Similarly, in the BRCA1 \emph{vs} BRCA2 network,
CDK1 and CKS1B belonged to a cluster that showed an increase for all its interactions.
These observations suggest that identifying clusters (complexes) that show
increase in co-expression could identify new therapeutic targets in cancer.


\section*{Conclusion}
Proteins seldom act in isolation, but instead interact to constitute specialized complexes driving key processes.
We integrated PPI and gene-expression datasets to perform a large-scale unbaised evaluation of complexes in PDAC and familial breast tumours.
These complexes showed considerable changes in expression, in particular decreases and countering increases,
reflecting compensatory processes coming into play in the tumours. These complexes enable us to explain the possible underlying mechanisms, 
which is otherwise difficult only by analysing individual genes.
These complexes are driven by the concerted action of influential TFs, which themselves work in cooperative and counteractive ways.
Network-based analysis shows that complexes could have therapeutic potential in cancer.



\section*{Methods}

The workflow for our computational approach is depicted in Figure 8, building on our earlier work~\cite{Srihari2013}.

\emph{From earlier work~\cite{Srihari2013} (upper portion of Figure 8):}
We first assemble a high-confidence network of human PPIs to identify human protein complexes.
These PPIs are largely devoid of contextual (conditional) information, and therefore we
overlay mRNA expression data of the coding genes, assigning a confidence score to each protein pair under normal and tumour conditions.
These scores reflect the presence or absence of interactions under these conditions.
Complexes are extracted from these conditional PPI networks by network clustering; for a detailed background on PPI networks and the complex-extraction procedure, see~\cite{Liu2009,Srihari2010,Srihari2012b,Kang2010}.

\emph{In this work (lower portion of Figure 8):} 
The contribution of this work is to relate changes in co-expression of complexes to changes in their transcriptional regulation by TFs between cancer conditions by introducing a \emph{log-linear model}. This enables us to identify influential TFs and to validate their roles in cancer. This procedure is described in the following subsections.

\subsection*{Measuring changes in co-expression of complexes between conditions}
Let $H=(V,E)$ be the human PPI network, where $V$ is the set of proteins and $E$ is the set of interactions among these proteins, and
$\mathcal{S} = \{S_1, S_2, .., S_n\}$ and $\mathcal{T} = \{T_1, T_2, ..., T_m\}$ be the sets of protein complexes extracted from $H$ under any two conditions, say normal and tumour.
For each complex $S_s \in \mathcal{S}$, we calculate its \emph{co-expression} as
\begin{equation}
C(S_s) = \frac{ \sum_{p,q \in S_s} \rho(p,q) }{ \binom{|S_s|}{2} },
\end{equation}
where $\rho(p,q)$ is the Pearson correlation for the protein pair $(p,q)$.
The $\rho$-values are Fisher-transformed, given by $z = \frac{1}{2}\ln(\frac{1+\rho}{1-\rho})$, which emphasizes the extreme $\rho$-values; 
for example, if $\rho$ = +/-$0.10$ then $z$ = +/-$0.043$, but if $\rho$ = +/-$0.99$ then $z$ = +/-$1.149$.
The co-expression values for $\mathcal{T}$ are calculated similarly.

To identify complexes that have changed co-expression between the conditions, we 
construct the set $\mathcal{M}(\mathcal{S},\mathcal{T})$ of \emph{matching complex pairs} such that every pair $(S_s,T_t) \in \mathcal{M}(\mathcal{S},\mathcal{T})$ satisfies
(a) a \emph{differential co-expression} $\Delta C(S_s, T_t) > 0$, and
(b) a minimum Jaccard similarity in protein composition $J(S_s,T_t) = \frac{|S_s \cap T_t|}{|S_s \cup T_t|} \geq t_J$,
where
\begin{equation}
\label{eq:deltaC}
\Delta C(S_s,T_t) = |C(S_s) - C(T_t)| \geq \delta > 0.
\end{equation}

We expect complexes disrupted between the two conditions to have changed their co-expression (including complete dissolution or new formation)
or have gained or lost a few proteins (rewiring within complexes) and therefore we use a $\delta > 0$ and a high $t_J$.

\subsection*{Relating changes in co-expression to changes in transcriptional regulation}

Let $\mathcal{F} = \{F_1, F_2, ..., F_k\}$ be the set of TFs. The \emph{regulation} by a TF $F_f \in \mathcal{F}$ of a complex $S_s$ is measured as
\begin{equation}
\label{eq:infl1}
R(S_s,F_f) = \frac{ \sum_{\{p: p \in S_s, (p,F_f) \in E\}} ~~\rho(p,F_f) }{|\{p: p \in S_s, (p,F_f) \in E \}|}, ~~~|\{p: p \in S_s, (p,F_f) \in E \}| > 0,
\end{equation}
where $\{p: p \in S_s, (p,F_f) \in E \}$ is the set of proteins of $S_s$ with which $F_f$ physically interacts in the network $H$.
The regulation by $F_f$ of the complexes $\mathcal{T}$ is measured similarly.

Here we consider a TF to regulate a complex only if the TF physically interacts (in the PPI network) with at least one protein in the complex.
From the classical view of transcriptional regulation, this assumption means that we consider a TF to regulate a set of genes encoding a complex only
if the TF physically interacts with at least one protein from that complex.
Although this assumption may be valid for only a subset of TFs or complexes, we employ it here to simplify our model.
Indeed (see Results) we only lose at most 13\% TF-complex pairs due to this assumption.

We then relate changes in regulation by TFs to changes in co-expression of complexes for $(S_s, T_t) \in \mathcal{M}(\mathcal{S},\mathcal{T})$ between the two conditions using a log-linear model
\begin{equation}
\label{eq:infl2}
{\Delta C(S_s,T_t) = \prod^{k}_{f=1} \Big(\Delta R(S_s,T_t,F_f) \Big)^{\gamma_f},}
\end{equation}
where $\Delta R(S_s,T_t,F_f) = |R(S_s,F_f) - R(T_t,F_f)|$ is the \emph{differential regulation} of the complex-pair $(S_s,T_t)$ by $F_f$, and
$\gamma_{f}$ is the \emph{influence coefficient} of $F_f$ in regulating the change $\Delta C(S_s,T_t)$.
Log-linear models are widely used to approximate non-linear systems because
they inherit the benefits of linear models yet allow a restricted non-linear relationship between inputs and outputs~\cite{LiaoJC2003}.

Equation~\ref{eq:infl2} can be written in matrix form after taking the logarithm as
\begin{equation}
\label{eq:infl3}
{\log[\Delta C(\mathcal{S},\mathcal{T})] = [\Gamma] \cdot \log[\Delta R(\mathcal{S},\mathcal{T},\mathcal{F})],}
\end{equation}
where $\log[\Delta C(\mathcal{S},\mathcal{T})]$ is a $|\mathcal{M}(\mathcal{S},\mathcal{T})| \times 1$ matrix of (log) differential co-expression of complexes,
$\log[\Delta R(\mathcal{S},\mathcal{T},\mathcal{F})]$ is a $|\mathcal{M}(\mathcal{S},\mathcal{T})| \times k$ matrix of (log) differential regulation by the TFs (here, $|\mathcal{M}(\mathcal{S},\mathcal{T})| > k$), 
and $[\Gamma]$ is a $k \times 1$ matrix of influence coefficients for the TFs.
Given this combinatorial regulation model, our purpose is to compute the influence coefficients $\gamma_f$ $(1 \leq f \leq k)$ by solving Equation~\ref{eq:infl3}, and for this we employ
singular-value decomposition (SVD) arriving at a least-squares solution~\cite{Yeung2002}.  
The TF displaying the highest (absolute) coefficient $|\gamma|$ has the highest overall influence in regulating changes in co-expression of complexes.

\subsubsection*{A simplified example to demonstrate our model}
Solving the equation can give positive as well as negative $\gamma$ values.
The absolute value $|\gamma|$ indicates the magnitude of the influence, while the sign indicates the direction: TFs of the same sign regulating a set of complexes work \emph{cooperatively},
while those of opposite signs work \emph{counteractively} with each other.
To understand this consider the following simplified example in which two TFs with influences $\gamma_1$ and $\gamma_2$ regulate two complexes $A$ and $B$ as per the following set of equations:
\begin{equation}
\begin{split}
\begin{aligned}
0.50 = (5)^{\gamma_1} \cdot (10)^{\gamma_2} ---- A\\
0.60 = (6)^{\gamma_1} \cdot (20)^{\gamma_2} ---- B,
\end{aligned}
\end{split}
\end{equation}
which after taking $\log_{10}$ becomes,
\begin{equation}
\begin{split}
\begin{aligned}
-0.301 = {\gamma_1} \cdot (0.699) + {\gamma_2} \cdot (1) ---- A\\
-0.221 = {\gamma_1} \cdot (0.778) + {\gamma_2} \cdot (1.301) ---- B.
\end{aligned}
\end{split}
\end{equation}
Here we see that the second TF performs at least twice the regulation than the first TF on the two complexes (5 and 6 \emph{vs} 10 and 20), the regulation by the second TF is
doubled (from 10 to 20) as against a smaller increase for the first TF (from 5 to 6) between $A$ and $B$, and yet $A$ and $B$ show roughly the same change in co-expression (0.50 \emph{vs} 0.60).
This intuitively means that the first TF has a greater influence than the second TF, and that counteracts the second TF to maintain the co-expression of complexes similar.
Indeed upon solving the equations we get $\gamma_1 = -1.293$ and $\gamma_2 = 0.603$, which is interpreted as the first TF being about twice as influential as the second, with
the two TFs working counteractively in regulating $A$ and $B$. It is easy to realize a similar example for the cooperative action of TFs.

\section*{Availability of supporting data}
The datasets used in this study are available at:
\url{http://www.bioinformatics.org.au/tools-data} under\\ CONTOUR\emph{v}2

\section*{Authors' contributions}
SS1 designed the study, performed the experiments and analysis, and wrote and revised the manuscript. KK helped in reviewing the analysis and providing biological interpretation for the results.
PBM, SS2, CL and PTS helped in data collection and analysis. MAR and KK supervised the project. All authors have read and approved the manuscript.

\section*{Acknowledgements}
    \emph{Funding}: Australian National Health and Medical Research Council (NHMRC) project grant 1028742 to PTS and MAR. PTS is supported by a fellowship from the National Breast Cancer Foundation Australia.


\newpage
{\ifthenelse{\boolean{publ}}{\footnotesize}{\small}
  \bibliography{bmc_article} }     


\ifthenelse{\boolean{publ}}{\end{multicols}}{}

\clearpage



\section*{Figures}

  \subsection*{Figure 1 - Changes in complex co-expression and its transcriptional regulation in pancreatic tumour}
   The complex CHUK-ECR1-IKBKB-IKBKG showed considerable change in its co-expression between normal and pancreatic tumour conditions. Its regulation by the NFKB family of TFs also
   exhibited similar changes. Both the complex and the TF family have been implicated in pancreatic tumours~\cite{Jones2008,Fujioka2003}.

  \subsection*{Figure 2 - Co-expression-wise distribution of PPIs in normal-PDAC and BRCA1-BRCA2 tumours}
  PPIs from both normal \emph{vs} PDAC and BRCA1 \emph{vs} BRCA2 conditions showed significant differences in their distributions of co-expression values.

  \subsection*{Figure 3 - Co-expression-wise distribution of complexes in normal-PDAC and BRCA1-BRCA2 tumours}
  Complexes from both normal \emph{vs} PDAC and BRCA1 \emph{vs} BRCA2 conditions showed significant differences in their co-expression distributions.    
  Note: a complex with negative co-expression for a condition possibly means that the complex does not exist under that condition.

  \subsection*{Figure 4 - Comparison of co-expression of complexes in normal-PDAC and BRCA1-BRCA2 tumours}
  Complexes from the two conditions were matched using $t_J = 0.67$ and $\delta = 0.10$ (see Methods) and differences between the maximum, average and minimum co-expression values
  were computed.
  
  The co-expression values are Fisher-transformed. A complex with negative co-expression for a condition possibly means non-existance of the complex under that condition.

   \subsection*{Figure 5 - Examples of dysfunctional complexes in pancreatic and breast tumours}
	Complexes showing changes in co-expression for (a) normal \emph{vs} PDAC, and
	(b) BRCA1 \emph{vs} BRCA2 tumours.

  \subsection*{Figure 6 - Fraction of known cancer genes constituting complexes}    
  \begin{itemize}
      \item Total COSMIC genes: 118 (present in the PPI network).      
      \item 66.1\% of known cancer genes in PDAC were covered in complexes (\#256) showing $\Delta C \geq 0.10$ between normal to PDAC tumour.
      \item 53.4\% of cancer genes in breast cancer were covered in complexes (\#277) showing $\Delta \geq 0.10$ between BRCA1 tumours and BRCA2 tumours.     
   \end{itemize}

  \subsection*{Figure 7 -- Differential PPI network constructed for normal \emph{vs} PDAC}
  (a) Largest component of the network composed of 519 interactions among 558 proteins, out of which 439 (or 84.5\%) interactions showed a decrease in co-expression from normal to PDAC   (red interactions indicate increase from normal to PDAC while green indicate decrease);
  (b) several genes are silenced by ubiquitination and sumoylation in PDAC;
  (c) several known drug targets are centered around interactions showing increase in co-expression.

 \subsection*{Figure 8 - Workflow of our approach}
  A log-linear model (lower portion of the workflow) is devised to relate changes in the co-expression of complexes with changes in their transcriptional regulation by TFs.

\clearpage


\section*{Tables}

  \subsection*{Table 1 - Complexes generated under pancreatic (normal-PDAC) and breast (BRCA1-BRCA2) conditions}
  The matched complexes $\mathcal{M}$ are generated by matching complexes between conditions using $t_J = 0.67 (= 2/3)$ and $\delta = 0.10$ (see Methods).

\begin{table}[ht]
\label{Tab:PPI_complexes}
{
\begin{tabular}{ l || c | c | c || l || c | c | c }
\hline																
						 \multicolumn{4}{c ||}{\em Pancreatic}	     								& 	\multicolumn{4}{c}{\em Breast}						\\
\hline
{}					&	\multicolumn{3}{c ||}{Complexes}							&{}			&	\multicolumn{3}{c}{Complexes}						\\	
{Condition}				&	{\#}		& {$|\mathcal{M}|$}		&	{\em Avg size}		&{Condition}		&	{\#}	& {$|\mathcal{M}|$}		& {\em Avg size}		\\
\hline
Normal					&	582		&				&				& BRCA1			&	547	&				&				\\
					&			&	256			&	8.20			&			&		&	277			&	7.87			\\
PDAC					&	581		&				&				& BRCA2			&	557	&				&				\\

\hline
\end{tabular}
 {}
 }    
\end{table}

\subsection*{Table 2 - Top enriched terms in KEGG pathways and Biological Process in disrupted complexes in pancreatic (normal \emph{vs} PDAC) and breast (BRCA1 \emph{vs} BRCA2) tumours (using DAVID~\cite{Dennis2003})}

\begin{table}[ht]
\label{Tab:Functional_enrichment1}
{\small
\begin{tabular}{l || l | c | l || l | c | l }
\hline
		&	\multicolumn{3}{c ||}{\em Pancreatic}							& \multicolumn{3}{c}{\em Breast}						\\	
\cline{2-7}
		&					& \multicolumn{2}{c ||}{Enrichment}			&					& \multicolumn{2}{c}{Enrichment}		\\
{Category}	&	{Annotation 	}	  	&{\em Genes (\%)}		& 	{\em $p$-value}		& {Annotation 	}			&{\em Genes(\%)}	& {\em $p$-value}	\\
\hline
		&					&				&				&					&			&			\\
		&	Cell cycle			&	4.6			&	3.5x$10^{-13}$		&	Cell cycle			&	3.2		&	2.7x$10^{-7}$	\\
		&	Pathways in cancer		&	6.2			&	5.8x$10^{-7}$		&	Pathways in cancer		&	5.8		&	2.9x$10^{-7}$	\\
		&	RIG-I-like receptor signalling  &	2.2			&	1.1x$10^{-5}$		&	Nucleotide excision repair	&	1.6		&	1.5x$10^{-5}$	\\
KEGG		&	Neurotrophin signalling		&	3.0			&	1.7x$10^{-5}$		&	DNA replication			&	1.4		&	6.4x$10^{-5}$	\\
pathways	&	Nucleotide excision repair	&	1.7			&	1.9x$10^{-5}$		&	Adipocytokine signalling	&	1.8		&	7.5x$10^{-7}$	\\
		&	Pancreatic cancer		&	2.1			&	5.7x$10^{-5}$		&	Apoptosis			&	2.1		&	1.2x$10^{-4}$  \\
		&	Adipocytokine signalling 	&	2.0			&	9.7x$10^{-5}$		&	Neurotrophin signaling		&	2.3		&	9.5x$10^{-4}$	\\
		&	Regulation of autophagy 	&	1.3			&	3.4x$10^{-4}$		&	Homologous recombination	&	1.0		&	1.6x$10^{-3}$	\\
		&	Mismatch repair	 		&	1.0			&	5.2x$10^{-4}$		&	Insulin signaling		&	2.2		&	6.0x$10^{-3}$	\\		
		&	Wnt signalling 			&	2.8			&	2.2x$10^{-3}$		&	Mismatch repair			&	0.9		&	2.8x$10^{-3}$	\\		
\hline
		&					&				&				&					&			&			\\
		&	Cell cycle			&	17.3			&	1.6x$10^{-35}$		&	Chromosome organization		&	14.3		&	1.5x$10^{-43}$	\\
Biological	&	Chromosome organization		&	13.0			&	6.2x$10^{-33}$		&	Chromatin organization		&	12.2		&	1.3x$10^{-40}$	\\
Process		&	RNA splicing			&	9.2			&	2.5x$10^{-28}$		&	Cell cycle			&	14.5		&	7.0x$10^{-25}$	\\		
		&	Chromatin modification		&	8.9			&	1.0x$10^{-27}$		&	Regulation of transcription	&	31.6		&	1.1x$10^{-24}$	\\
\hline
\multicolumn{7}{l}{\small $\sim 700$ genes tested in pancreatic and $\sim 550$ genes tested in breast}\\

\end{tabular}
}
\end{table}

\clearpage

\subsection*{Table 3 - Changes in co-expression of complexes in pancreatic (normal-PDAC) and breast (BRCA1-BRCA2) conditions}
(The co-expression values are Fisher-transformed)

\begin{itemize}
\item \noindent 
$\Delta C$ represents the (absolute) change in co-expression of complexes between conditions, that is, increase or decrease in co-expression of complexes from normal to PDAC or BRCA1 to BRCA2 tumours.

\item
$\mathcal{M'}$ -- complexes with co-expression for normal $>$ PDAC tumour or BRCA1 $>$ BRCA2 tumour\\
$\mathcal{M''}$ -- complexes with co-expression for PDAC tumour $>$ normal or BRCA2 $>$ BRCA1 tumour.
\end{itemize}

\begin{table}[ht]
\label{Tab:complex_correlation_changes}
{
\begin{tabular}{l || c || c || c | c || c || c | c  }
\hline						   

			&									& \multicolumn{3}{c ||}{\em Pancreatic}				&			\multicolumn{3}{c}{\em Breast}			\\
\cline{3-8}
{Complex}		&	{}			  					&{Size of}		&\multicolumn{2}{c ||}{$\Delta C$}	&{Size of}		&\multicolumn{2}{c}{$\Delta C$}	  		\\
{type}			&	{Subset}	  						&{subset}		& {\em Max}	&{\em Avg}		&{subset}		& {\em Max}	&{\em Avg}			\\
\hline
			&	 $\mathcal{M}'$							& 159			& 0.969		& 0.336 		& 225			& 0.761		& 0.281 			\\
Our
			&	 $\mathcal{M}''$						& 96			& 0.421		& 0.192 		& 52			& 0.543		& 0.197 			\\
\hline
			&	$\mathcal{M}'$							& 138			& 0.602		& 0.219			& 386			& 0.642		& 0.178				\\
CORUM
			&	$\mathcal{M}''$							& 51			& 0.512		& 0.209			& 55			& 0.347		& 0.212				\\
%

\hline

\end{tabular}																												
}
\end{table}

\subsection*{Table 4 - Relationship between changes in co-expression of complexes ($\Delta C$) and changes in transcriptional regulation by TFs ($\sum \Delta R$)}

All coefficient values are positive, and Spearman coefficients were higher than Pearson, indicating a non-linear relationship between $\Delta C$ and $\sum \Delta R$.

\begin{table}[ht]
\label{Tab:complex_correlation_changes}
{
\begin{tabular}{l || c | c | c || c | c | c }
\hline						   
		&	\multicolumn{3}{c ||}{\em Pancreatic}		&	\multicolumn{3}{c}{\em Breast}				\\
\cline{2-7}
{Complex}	&	{\#Complex-TF}	& {}		& {}		&	{\#Complex-TF}	 & {}			&  {}		\\
{type}		&	{pairs}		& {Pearson}	&{Spearman}	&	{pairs}		 & {Pearson}		& {Spearman}	\\
\hline
		&			& 		&		&			 &			&		\\
Our		&	226		& 0.273		& 0.436		&	241		 &	0.206		&	0.413	\\
\hline
		&			&		&		&			 &			&		\\		
CORUM		&	71		& 0.298		& 0.434		&	32		 &	0.212		&	0.393	\\
\hline
\end{tabular}																									
}
\end{table}

\clearpage

\subsection*{Table 5 - Influential TFs identified in pancreatic (normal \emph{vs} PDAC) and breast (BRCA1 \emph{vs} BRCA2) tumours}
TFs are ranked based on their overall influence $(|\gamma|)$ for normal \emph{vs} PDAC and BRCA1 \emph{vs} BRCA2 tumours.
The normalized-$\gamma$ values are obtained by dividing the $\gamma$ values by the maximum $|\gamma|$.
The significance ($p$ value) was computed against 10000 background influence values, each generated from 10000 random shuffles of gene symbols in the expression dataset.
The adjusted $p$-values for differential expression (DE) analysis -- whether up- ($\uparrow$) or down- ($\downarrow$) regulated from normal to PDAC
and from BRCA1 to BRCA2 tumours -- were computed using limma~\cite{Smyth2004}.

\begin{table}[ht]
{\scriptsize
\begin{tabular}{ l | c | c | c | l || l || l | c | c || c || l | l }

\hline	
			\multicolumn{6}{c ||}{\em Pancreatic}										& \multicolumn{6}{c }{\em Breast}																						\\
\hline																																											
				&				&		& {\#Complexes}		&		& {DE}							&			&			&			& {\#Complexes}		&			& {DE}						\\
{TF}				&	 {$\gamma$}		& {Norm-$\gamma$}& {regulated}		&{$p$-val}	& {($p$-val)}						&	{TF}		& {$\gamma$}		& {Norm-$\gamma$}	& {regulated}		&{$p$-val}		& {($p$-val)}					\\
\hline		
STAT5B				&	2.911			&1.000		&	188		& 0.001		& 0.021$\downarrow$					&	TBP		&	-5.153		&	-1.000		&	199		& 0.001			& NS						\\
RREB1				&	-2.606			&-0.895		&	190		& 0.004		& NS							&	BACH1		&	4.221		&	0.819		&	199		& 0.003			& NS						\\
BACH1				&	-1.592			&-0.547		&	188		& 0.001		& 5.48E-05$\uparrow$					&	POU3F2		&	-3.902		&	-0.757		&	199		& 0.007			& NS						\\
SRF				&	1.522			&0.523		&	40		& 0.004		& 3.06E-05$\downarrow$					&	RREB1		&	2.654		&	0.515		&	200		& 0.002			& NS						\\
IRF2				&	-1.462			&-0.502		&	37		& 0.014		& 1.24E-06$\downarrow$					&	ATF6		&	-1.185		&	-0.230		&	201		& 0.002			& NS						\\
AHR				&	1.359			&0.467		&	190		& 0.003		& 1.93E-11$\uparrow$					&	SRF		&	1.079		&	0.209		&	40		& 0.007			& NS						\\
TBP				&	-0.944			&-0.324		&	188		& 0.002		& 0.001$\uparrow$					&	STAT5B		&	1.022		&	0.198		&	199		& 0.007			& NS						\\
POU3F2				&	-0.731			&-0.251		&	188		& 0.002		& 6.64E-07$\downarrow$					&	SIN3A		&	0.980		&	0.190		&	205		& 0.007			& NS						\\
IRF7				&	-0.654			&-0.225		&	188		& 0.002		& 2.50E-06$\uparrow$					&	TCF4		&	0.601		&	0.117		&	40		& 0.007			& NS						\\
HSF1				&	-0.621			&-0.213		&	190		& 0.001		& NS							&	TAL1		&	0.599		&	0.116		&	14		& 0.007			& NS						\\
YY1				&	0.619			&0.213		&	195		& 0.001		& 3.80E-05$\uparrow$					&	CEBP$\beta$	&	-0.537		&	-0.104		&	201		& 0.002			& NS						\\
STAT1				&	0.568			&0.195		&	190		& 0.005		& 4.79E-06$\uparrow$					&	SOX9		&	0.447		&	0.087		&	204		& 0.008			& NS						\\
TP53				&	0.543			&0.187		&	190		& 0.004		& 6.34E-06$\uparrow$					&	GATA3		&	0.422		&	0.082		&	90		& 0.001			& 0.012$\uparrow$				\\
XBP1				&	-0.549			&-0.189		&	190		& 0.004		& 0.002$\downarrow$					&	STAT1		&	0.411		&	0.080		&	202		& 0.001			& NS						\\
SOX9				&	0.504			&0.173		&	193		& 0.004		& 2.43E-05$\uparrow$					&	STAT5A		&	0.407		&	0.079		&	200		& 0.004			& NS						\\
CEBP$\beta$			&	0.427			&0.147		&	191		& 0.005		& 1.71E-08$\uparrow$					&	NF$\kappa$B1	&	0.390		&	0.076		&	202		& 0.001			& NS						\\
ATF6				&	-0.419			&-0.144		&	191		& 0.008		& NS							&	IRF2		&	-0.389		&	-0.075		&	42		& 0.006			& NS						\\
STAT5A				&	0.385			&0.132		&	188		& 0.001		& 0.004$\uparrow$					&	GATA2		&	-0.323		&	-0.063		&	40		& 0.001			& NS						\\
MSX1				&	0.366			&0.126		&	188		& 0.002		& 0.008$\uparrow$					&	FOXO1		&	0.305		&	0.059		&	30		& 0.001			& NS						\\
GATA2				&	0.328			&0.113		&	40		& 0.001		& 9.96E-03$\downarrow$					&	ESR1		&	0.282		&	0.055		&	24		& 0.001			& 0.008$\uparrow$				\\
TAL1				&	0.319			&0.110		&	12		& 0.003		& 1.09E-07$\downarrow$					&	PPAR$\gamma$	&	0.279		&	0.054		&	60		& 0.001			& NS						\\
FOXO1				&	-0.306			&-0.105		&	30		& 0.003		& 0.001$\uparrow$					&	TP53		&	0.273		&	0.053		&	206		& NS			& NS						\\
ESR1				&	0.254			&0.087		&	27		& 0.004		& 3.43E-04$\uparrow$					&	YY1		&	0.274		&	0.053		&	206		& 0.009			& NS						\\
JUNB				&	0.219			&0.075		&	198		& 0.004		& 2.44E-07$\uparrow$					&	STAT3		&	-0.222		&	-0.043		&	200		& 0.004			& NS						\\
CEBP$\alpha$			&	0.187			&0.064		&	194		& 0.001		& NS							&	MYB		&	-0.213		&	-0.041		&	60		& 0.003			& NS						\\
STAT3				&	0.169			&0.058		&	190		& 0.004		& 2.34E-06$\uparrow$					&	GATA1		&	0.204		&	0.040		&	40		& 0.001			& NS						\\
SP1				&	0.159			&0.055		&	10		& 0.001		& 0.003$\uparrow$					&	FOXA1		&	0.173		&	0.034		&	207		& 0.001			& 0.006$\uparrow$				\\
TCF3				&	0.155			&0.053		&	40		& 0.001		& NS							&	MSX1		&	0.170		&	0.033		&	199		& NS			& NS						\\
BRCA1				&	0.146			&0.050		&	200		& 0.003		& 0.003$\uparrow$					&	HSF2		&	-0.165		&	-0.032		&	201		& NS			& NS						\\
SIN3A				&	0.138			&0.047		&	193		& 0.001		& 2.05E-4$\uparrow$					&	FOXO3		&	0.147		&	0.029		&	50		& 0.009			& NS						\\
PPAR$\gamma$			&	0.124			&0.043		&	13		& 0.001		& NS							&	PPAR$\alpha$	&	0.133		&	0.026		&	40		& NS			& NS						\\
HSF2				&	0.120			&0.041		&	191		& 0.001		& 0.004$\uparrow$					&	FOS		&	0.121		&	0.023		&	44		& 0.005			& NS						\\
MAX				&	0.113			&0.039		&	80		& NS& N		S							&	SP1		&	0.098		&	0.019		&	18		& 0.003			& NS						\\
NF$\kappa$B1			&	0.106			&0.036		&	192		& 0.002		& NS							&	TCF3		&	-0.096		&	-0.019		&	90		& 0.009			& NS						\\
FOS				&	0.044			&0.015		&	42		& 0.001		& NS							&	CEBP$\alpha$	&	0.094		&	0.018		&	202		& 0.009			& NS						\\
MYB				&	-0.023			&-0.008		&	40		& 0.002		& NS							&	NF$\kappa$B2	&	-0.091		&	-0.018		&	202		& 0.003			& NS						\\
NF$\kappa$B2			&	0.018			&0.006		&	193		& 0.002		& NS							&	JUN		&	0.080		&	0.016		&	205		& 0.003			& NS						\\
FOXO3				&	-0.001			&0.000		&	90		& 0.004		& NS							&	AHR		&	0.076		&	0.015		&	200		& NS			& NS						\\
--				&	--			&		&	--		&		&							&	XBP1		&	0.059		&	0.011		&	201		& 0.002			& 0.017$\uparrow$				\\
--				&	--			&		&	--		&		&							&	HSF1		&	-0.056		&	-0.011		&	200		& 0.001			& NS						\\
--				&	--			&		&	--		&		&							&	IRF7		&	-0.033		&	-0.006		&	199		& 0.009			& NS						\\

\hline
\multicolumn{8}{l}{NS: Not significant at $p < 0.05$. \#Genes tested in each case: $\sim20700$.}\\
\end{tabular}																												
}
\end{table}


%

\end{bmcformat}
\end{document}